# Saturation of ion irradiation effects in Cr$_2$AlC

Qing Huang[a], Han Han[a], Renduo Liu[a], Guanhong Lei[a], Long Yan[a*], Jie Zhou[b], Qing Huang[b]

[a]Shanghai Institute of Applied Physics, Chinese Academy of Sciences (CAS), Shanghai 201800, China.

[b]Ningbo Institute of Material Technology & Engineering, Chinese Academy of Sciences (CAS), Ningbo 315201, China.

*Corresponding author. Tel.: +86-21-39194773; Fax: +86-21-39194539. E-mail address: yanlong@sinap.ac.cn

**Abstract**

Cr$_2$AlC materials were irradiated with 7 MeV Xe$^{26+}$ ions and 500 keV He$^{2+}$ ions at room temperature. A structural transition with an increased $c$ lattice parameter and a decreased $a$ lattice parameter occurs after irradiation to doses above 1 dpa. Nevertheless, the modified structure is stable up to the dose of 5.2 dpa without obvious lattice disorder. The three samples irradiated to doses above 1 dpa have comparable lattice parameters and hardness values, suggesting a saturation of irradiation effects in Cr$_2$AlC. The structural transition and irradiation effects saturation are ascribed to irradiation-induced antisite defects (Cr$_{Al}$ and Al$_{Cr}$) and C interstitials, which is supported by the calculations of the formation energies of various defects in Cr$_2$AlC. The irradiation-induced antisite defects and C interstitials may be critical to understand the excellent resistance to irradiation-induced amorphization of MAX phases.

## 1. Introduction

Cr$_2$AlC, which is the only ternary compound in the Cr-Al-C system, has a hexagonal crystal structure with a space group of P6$_3$/mmc, in which Cr$_2$C layers are interleaved with layers of Al. Cr$_2$AlC belongs to the family of layered ternary compounds known as M$_{n+1}$AX$_n$ (MAX) phases where n is 1, 2 or 3, M is an early transition metal, A is an A group element, X is C or N. MAX phases have attracted increasing attention since they offer a unique combination of the merits of both metals and ceramics [1]. Briefly,

Cr$_2$AlC is relatively soft (Vickers hardness of 3.5 - 5.5 GPa), elastically stiff (Young's modulus of 278 GPa, shear modulus of 116 GPa) and readily machinable [2,3].

The MAX phases have been proposed to be candidate materials for structural and fuel coating applications in the Generation IV Nuclear reactors [4-8]. It is of particular importance for reactor materials to investigate their tolerance to irradiation damage. Up to now, two 312 phases, Ti$_3$SiC$_2$ and Ti$_3$AlC$_2$, have been investigated on their response to irradiation using heavy ions and He ions [5-16]. After Xe-ion irradiation to a dose of 10 dpa in Ti$_3$SiC$_2$ [16] and He-ion irradiation to a dose of 31 dpa in Ti$_3$AlC$_2$ [15], the nanolamellar structure disappeared, but both materials remained crystalline, indicating excellent resistance to amorphization of the two 312 phases. Besides, irradiation induced phase transition from α phase to β phase has been reported in ion-irradiated Ti$_3$SiC$_2$ [9] and Ti$_3$AlC$_2$ [15]. By Rietveld refinement of the XRD patterns, we found that Ti$_3$AlC$_2$ experienced a severer phase transition than Ti$_3$SiC$_2$ [16].

To our knowledge, there is no ion irradiation effects reported for bulk 211 phases. Good high-temperature oxidation and hot corrosion resistance endow Cr$_2$AlC with the potential to be used in high-temperature corrosive environments, which encouraged us to expand ion irradiation applications to Cr$_2$AlC. In this study, Cr$_2$AlC samples were irradiated with 7 MeV Xe$^{26+}$ ions and 500 keV He$^{2+}$ ions at room temperature. The microstructure and hardness changes were characterized by TEM, XRD, and nano-indentation, respectively. The mechanism of irradiation-induced antisite defects and carbon interstitials, firstly proposed by Yang et al. [15], were applied to explain the evolutions of XRD patterns and SEAD patterns. The formation energies of various defects in Cr$_2$AlC were calculated in order to further verify the mechanism.

## 2. Experimental procedures

2.1. Material

The polycrystalline Cr$_2$AlC samples were synthesized by the Ningbo Institute of Material Technology and Engineering, Chinese Academy of Science (IMTE-CAS). Stoichiometric mixtures of commercial Cr, Al, C powders were sintered in a spark plasma sintering (SPS) furnace at 1500 ºC for 35 min, under a uniaxial pressure of 30 MPa. The as-received samples were sectioned and polished with fine metallographic abrasive paper with silicon carbide suspensions.

2.2. Irradiation

Ion irradiation experiments were carried out at room temperature in a terminal of the 320 kV High-voltage Experimental Platform equipped with an electron cyclotron resonance (ECR) ion source in the Institute of Modern Physics, Chinese Academy of Science (IMP-CAS). The ions were 7 MeV $Xe^{26+}$ ions ($4 \times 10^{14}$, $2 \times 10^{15}$ ions/cm$^2$) and 500 keV $He^{2+}$ ions ($1 \times 10^{16}$, $1 \times 10^{17}$ ions/cm$^2$). The ion irradiation processes were simulated by SRIM 2008 program using the "Kinchin-Pease quick calculation" mode with threshold displacement energies for each element being 25 - 28 eV [7,13]. The vacancy profiles (Fig. 1(a)) in $Cr_2AlC$ produced by 7 MeV $Xe^{26+}$ ions and 500 keV $He^{2+}$ ions were derived from the simulation and were used to calculate the dpa profiles (Fig. 1(b)). The dpa value is not uniform through the irradiation layer. The peak dpa value is 5.2 in Xe-ion-irradiated $Cr_2AlC$ at a dose of $2 \times 10^{15}$ ions/cm$^2$, and is 3 in He-ion-irradiated $Cr_2AlC$ at a dose of $1 \times 10^{17}$ ions/cm$^2$.

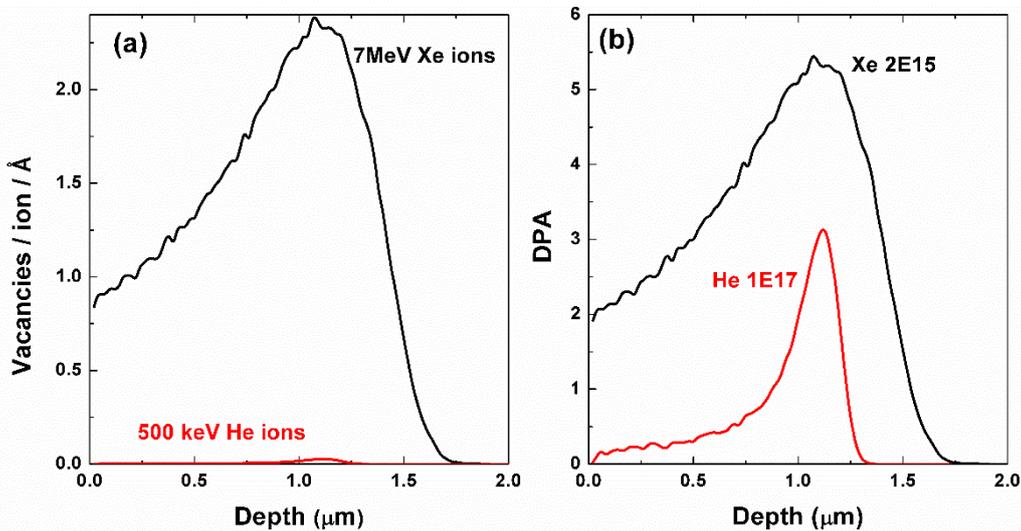

Fig. 1. (a) Vacancy profiles induced by 7 MeV $Xe^{26+}$ ions and 500 keV $He^{2+}$ ions in $Cr_2AlC$ material (SRIM calculation), and (b) corresponding dpa profiles at doses of $2 \times 10^{15}$ (Xe ion) and $1 \times 10^{17}$ (He ion) ions/cm$^2$.

2.3. Characterization techniques

The irradiated $Cr_2AlC$ samples were observed using a 200 kV Tecnai G2 F20 transmission electron microscope (TEM). The irradiation layers were examined by performing characterization on cross section specimens machined from the bulk samples. The specimens were prepared as follows: two small bars were cut from the irradiated bulk samples and the irradiated surfaces were joined face-to-face with glue. The edges were milled and the specimen (with glue around) was inserted into a copper tube (3 mm in diameter). After that, the tube was sliced into small pieces which then were mechanical milled down to around 50 μm in thickness. The middle of the thin foil was further thinned using a dimple grinder to about 20 μm. At last, 5-keV Ar-ion milling was used to obtain a penetration hole in the middle. The damage to the crystal lattice was analyzed using both bright-field imaging (BF) and selected-area electron diffraction (SAED).

The irradiated and un-irradiated $Cr_2AlC$ samples were characterized by low-incidence X-ray diffraction (LI-XRD) using a Bruker D8 Advance diffractometer equipped with copper anticathode (0.154 nm). The diffractograms were recorded between 10º and 70º in 2θ scale, under an incidence of 1º for all samples. All diffractograms were analyzed by the Rietveld refinement method using the Bruker TOPAS program. Pseudo-Voigt function was selected to refine peak profile.

A Berkovich diamond indenter tip with a radius of 20 nm was used to perform the nano-indentation measurements on the surfaces of the $Cr_2AlC$ samples. The maximum penetration depth was set at 1 μm. About 35 - 45 indents were measured for each sample. The distance between indentations was larger than 50 μm.

## 3. Computational details

The formation energies of various defects in $Cr_2AlC$ were calculated under the framework of density functional theory as implemented in VASP package [17]. Exchange and correlation effects were treated by the generalized gradient approximation proposed by Perdew et al [18]. Electron-ion interactions were described by the projector augmented plane-wave method [19], and the wave functions were expanded in a plane-wave basis set with an energy cutoff of 400 eV. Calculation of the defect structure employed a $3\times3\times1$ supercell, which contains 72 atoms. The special k-point sampling integration was used over the Brillouin zone by using the Monkhorst-Pack method with $4\times4\times2$ for this supercell [20]. The lattice parameters and internal freedom of the unit cell were fully optimized until the total energy difference was smaller than $1 \times 10^{-6}$ eV. According to our previous studies on defects in 211 phase $Ti_2AlC$ [21], the supercell has been proved to be big enough to reproduce the defect structures. In this calculation, the chemical potentials of species were obtained from the total energies of the bulk systems: body-centered cubic Cr metal, face-centered cubic Al metal and graphite.

## 4. Results

4.1. TEM characterization

$Cr_2AlC$ has a hexagonal structure with space group $P6_3/mmc$. Fig 2(a) shows the unit

cell of $Cr_2AlC$ crystal. The Wyckoff positions are 4f for Cr atoms, 2d for Al atoms, and 2a for C atoms [3]. The stacking sequence of $Cr_2AlC$ along the [0001] direction can be described as A<u>B</u>AB<u>A</u>B, where the underlined letters refer to Al layers and other letters refer to Cr layers. This stacking sequence can be seen in the high resolution TEM image (Fig. 2(b)), with the electron beam paralleling to the [11-20] direction. The corresponding SAED pattern is shown in Fig. 2(c). The lattice parameters were derived from the SAED pattern, and are $a$ = 0.286 nm and $c$ = 1.282 nm, which are in good agreement with the previous measurements [2].

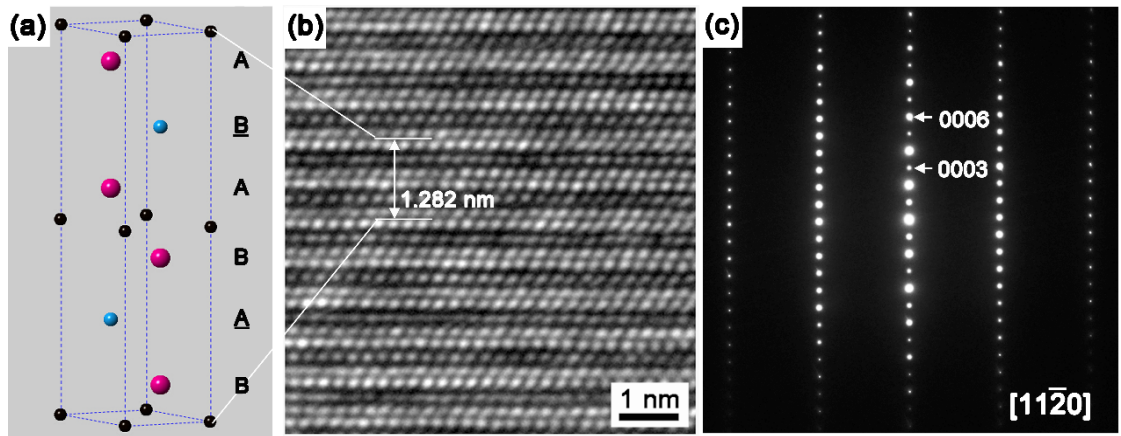

Fig. 2. (a) Unit cell of $Cr_2AlC$. The stacking sequence is illustrated. (b) High resolution TEM image and (c) corresponding SAED pattern of un-irradiated $Cr_2AlC$ with the electron beam parallel to the [11-20] direction.

After irradiation, the $Cr_2AlC$ sample irradiated with 7 MeV $Xe^{26+}$ ions at a dose of $2 \times 10^{15}$ ions/cm$^2$ was analyzed using TEM. A crystal grain astride the border line between the irradiated layer and the substrate was chosen for investigation. First, the un-irradiated part of the grain was observed, in order to make sure that the electron beam is parallel to the [11-20] direction. Then the irradiated part of the grain which is also located at the end of the irradiation layer (5.2 dpa) was characterized. High resolution TEM image and corresponding SAED pattern are shown in Fig. 3. The lattice is not damaged after irradiation, except the Cr layer and Al layer cannot be distinguished through phase contrast. A simple ABABAB sequence is observed in Fig. 3(a). Comparing the SAED patterns of irradiated and un-irradiated samples, many diffraction spots disappeared after irradiation. In the {000$l$} reflections, diffraction spots (0003) and (0006) exist while the diffraction spots (000$l$) ($l$ = 1, 2, 4, 5) totally disappear in Fig. 3(b). The lattice parameter $c$ (= 1.345 nm) derived from the SAED pattern is illustrated in Fig. 3(a), and is larger than that of the un-irradiated $Cr_2AlC$.

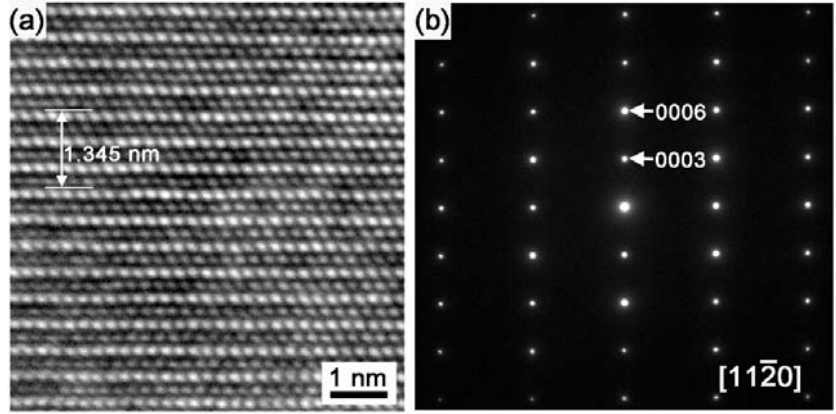

Fig. 3. (a) High resolution TEM image and (b) corresponding SAED pattern of $Cr_2AlC$ sample irradiated with $Xe^{26+}$ ions at a dose of $2 \times 10^{15}$ ions/cm$^2$. The electron beam is parallel to the [11-20] direction.

The decrease in the amount of diffraction spots can be attributed to the increase in symmetry. The indistinguishability between the Cr layer and Al layer after irradiation can be explained based on a mechanism: irradiation produced many antisite defects, $Cr_{Al}$ and $Al_{Cr}$, leading to a mixture of Cr and Al atoms in each layer. To simulate this irradiation effect, the Al atoms in the structure shown in Fig. 2(a) was replaced by Cr atoms. The electron diffraction pattern of the modified structure was calculated by using CrystalMaker software. The diffraction spots (0002) and (0004) weakened (compared with diffraction spot (0006)), but did not disappear.

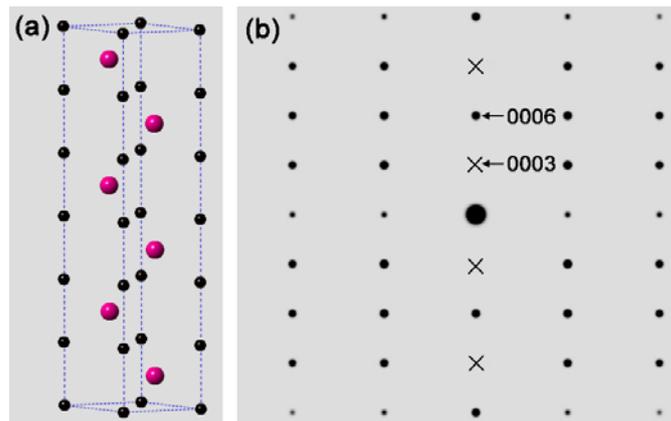

Fig. 4. (a) Modified unit cell based on the hypotheses of irradiation-induced antisites and C interstitials, and (b) corresponding calculated electron diffraction pattern which is consistent with the SAED pattern in Fig. 3(b).

In the unit cell shown in Fig. 2(a), the space between the two adjacent Cr layers is larger than the space between the adjacent Cr layer and Al layer. To make the diffraction spots (0002) and (0004) disappear, the interlamellar spacing has to be uniform, which could be achieved based on a mechanism: ion irradiation produced many C interstitials randomly occupying the octahedral holes between the original Cr layer and Al layer. This mechanism would lead to an expansion of the lattice in the *c* direction, which is consistent with the measurement value of the lattice parameter *c*. To simulate this irradiation effect, the structure shown in Fig. 2(a) was further modified and is shown in Fig. 4(a). In the modified structure, the original Al atoms were replaced by Cr atoms, and every octahedral hole is occupied by a C atom, resulting in a uniform interlamellar spacing. The electron diffraction pattern of the modified structure (Fig. 4(b)) shows that the diffraction spot (0002) and (0004) disappeared, which is in good agreement with the SAED pattern after irradiation. The forbidden diffraction spot (0003) indicated by the cross in Fig. 4(b) appears in the SAED pattern, which can be attributed to double diffraction [3].

The other irradiated $Cr_2AlC$ samples were also characterized with TEM. The three samples irradiated with He ions at a dose of $1 \times 10^{17}$ ions/cm$^2$ and Xe ions at doses of $4 \times 10^{14}$ and $2 \times 10^{15}$ ions/cm$^2$ have similar SAED patterns (Fig. 3(b)). The peak dpa values in these three samples are larger than 1. The sample irradiated with He ions at a dose of $1 \times 10^{16}$ ions/cm$^2$ (0.3 dpa) showed a similar SEAD pattern to that of the un-irradiated sample (Fig. 2(c)). The lattice parameters were measured from the SAED patterns, and are shown in Table 1.

Table 1. lattice parameters of $Cr_2AlC$ samples derived from the SAED patterns

|  | Dose (ions/cm$^2$) | peak dpa | *a* (nm) | *c* (nm) |
|---|---|---|---|---|
| virgin | 0 | 0 | 0.2860 | 1.282 |
| He ions | $1 \times 10^{16}$ | 0.3 | 0.2848 | 1.290 |
|  | $1 \times 10^{17}$ | 3.0 | 0.2806 | 1.336 |
| Xe ions | $4 \times 10^{14}$ | 1.04 | 0.2802 | 1.339 |
|  | $2 \times 10^{15}$ | 5.2 | 0.2795 | 1.345 |

4.2. Low-incidence XRD

The XRD patterns of un-irradiated and irradiated $Cr_2AlC$ samples at low-incidence angle of 1° are shown in Figs. 5(a) - 5(e). Miller-Bravais indices of XRD peaks for the un-irradiated sample are shown in Fig. 5(a). The XRD pattern in Fig. 5(b) is similar to that of the un-irradiated sample, indicating that the material remained crystalline without obvious damage and structural transition after He-ion irradiation at a dose of $1 \times 10^{16}$ ions/cm$^2$. The three $Cr_2AlC$ samples irradiated to doses above 1 dpa show similar XRD patterns in Fig. 5(c) - 5(e). The peak (0002) disappeared and a new peak emerged at 40º in Fig. 5(c) - 5(e).

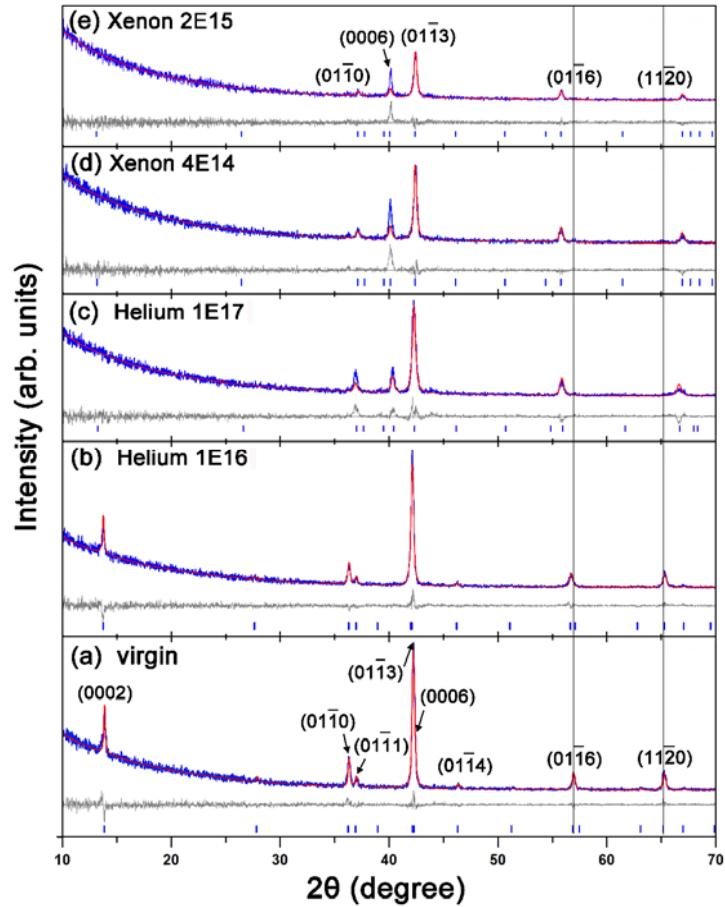

Fig. 5. XRD patterns of un-irradiated and irradiated $Cr_2AlC$ at low-incidence angle of 1º. Miller-Bravais indices of XRD peaks are shown in the figures. Rietveld refinement were performed using the original structure (Fig. 2(a)) for XRD patterns in (a) and (b), and the modified structure (Fig. 4(a)) for XRD patterns in (c) - (e).

Rietveld refinement method was used to analyze the XRD pattern evolution after irradiation. For the un-irradiated $Cr_2AlC$, the original $Cr_2AlC$ structure (Fig. 2(a)) was used in the Rietveld refinement. The atomic positions in the unit cell (Fig. 2(a)) are C (0, 0, 0), Cr (1/3, 2/3, 0.087), and Al (2/3, 1/3, 0.25). For the irradiated $Cr_2AlC$, the original $Cr_2AlC$ structure was used in the refinement of the XRD pattern shown in Fig. 5(b), while the modified $Cr_2AlC$ structure (Fig. 4(a)) was used in the refinements of the XRD patterns shown in Figs. 5(c) - 5(e). The simulated XRD patterns are in good agreement with the measurement data. The weighted reliability factors ($R_{wp}$) of these refinements and the lattice parameters are shown in Table 2.

Table 2. Rietveld refinement results of un-irradiated and irradiated $Cr_2AlC$ samples

|  | Dose (ions/cm$^2$) | peak dpa | Phase | $a$ (nm) | $c$ (nm) | $R_{wp}$ |
|---|---|---|---|---|---|---|
| Virgin | 0 | 0 | Original $Cr_2AlC$ | 0.2859 | 1.2814 | 14.320 |
| He ions | $1 \times 10^{16}$ | 0.3 | Original $Cr_2AlC$ | 0.2856 | 1.2895 | 15.056 |
|  | $1 \times 10^{17}$ | 3.0 | Modified $Cr_2AlC$ | 0.2802 | 1.3381 | 16.254 |
| Xe ions | $4 \times 10^{14}$ | 1.04 | Modified $Cr_2AlC$ | 0.2794 | 1.3479 | 16.292 |
|  | $2 \times 10^{15}$ | 5.2 | Modified $Cr_2AlC$ | 0.2794 | 1.3482 | 15.431 |

It was reported that $AlCr_2$, $Cr_7C_3$, and $Al_8Cr_5$ are the intermediate phases during the synthesis of $Cr_2AlC$ [3]. We also tried these three phases in the refinements of the XRD patterns shown in Figs. 5(c) - 5(e), but none of them matches these XRD patterns.

By Rietveld refinement analysis, the Miller-Bravais indices of XRD peaks for the modified crystal structure are indicated in Fig. 5(e). After irradiation to doses above 1 dpa, the peak (0006) shifted from 42.5° (before irradiation) to 40°, indicating an expansion of the lattice in the $c$ direction. Besides, the peak (01-16) shifted to the left, while the peak (11-20) shifted to the right after irradiation.

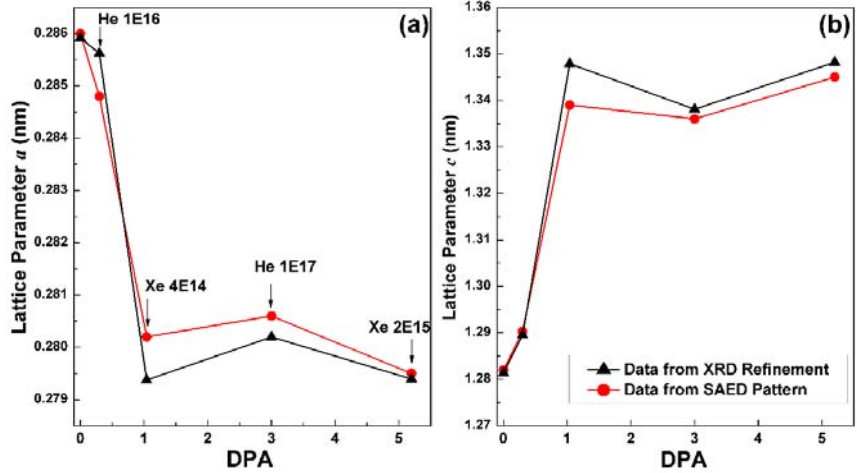

Fig. 6. (a) Lattice parameter $a$ and (b) lattice parameter $c$ versus the dpa value. The irradiation ions and doses corresponding to the dpa values are indicated in (a).

The lattice parameters achieved from the Rietveld refinements and those derived from the SAED patterns are shown in Figs 6(a) and 6(b) as a function of the peak dpa values derived from SRIM calculations (Fig. 1). The irradiation ions and doses corresponding to the dpa values are indicated in Fig. 6(a). Both methods show an increase of $c$ and a decrease of $a$ after irradiation.

4.3. Nano-indentation

Figure 7 shows the changes in hardness of $Cr_2AlC$ samples irradiated with Xe ions and He ions at various doses. The hardness value is depth dependent, because of an indentation size effect, wherein the hardness increases with decreasing indentation size, which has been well studied by Nix and Gao [22]. The hardness decreased with increasing indentation depth. But the nano-indentation test could not provide a measurement of the true hardness at each depth. The reason is that the stress field induced by the indenter extends much deeper than the contact depth [11].

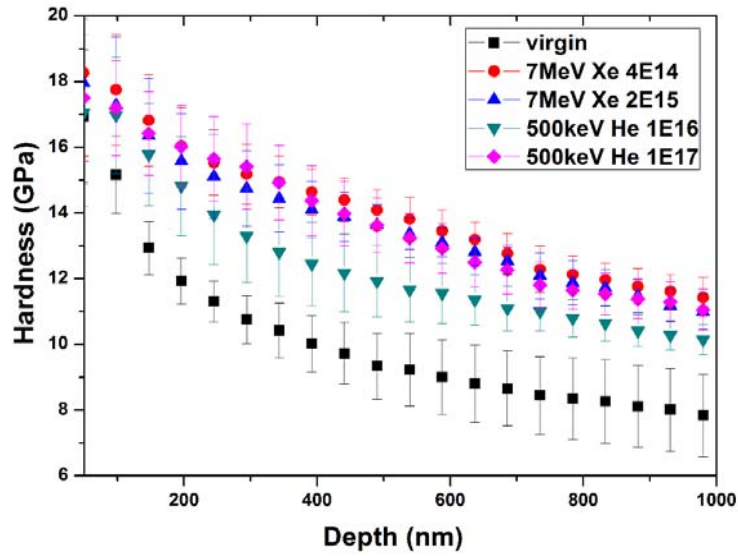

Fig. 7. Hardness versus indentation depth for un-irradiated and irradiated $Cr_2AlC$.

There is an evident hardening effect of ion irradiation in the ion-irradiated $Cr_2AlC$ samples. The hardness increases with He ion dose. The three samples irradiated to doses above 1 dpa show comparable hardness values.

4.4. Defect formation energy calculations

To verify the mechanism of irradiation-induced antisite defects ($Cr_{Al}$ and $Al_{Cr}$) and C interstitials, the formation energies of various defects in $Cr_2AlC$ were calculated. Firstly, the lattice parameters of perfect $Cr_2AlC$ were calculated and are $a$ = 0.2843 nm and $c$ = 1.2674 nm, which are slightly smaller than the measurement data. The formation energies of on-lattice defects in $Cr_2AlC$ including vacancies, interstitials, antisite defects are shown in Table 3. The interstitial position with the largest free volume is denoted by the octahedra located between the Cr layer and Al layer (Fig. 8), which is also the position that occupied by the added C atoms in the modified $Cr_2AlC$ structure (Fig. 4(a)). The antisite defects $C_{Cr}$, $Cr_C$, $C_{Al}$, and $Al_C$ were not included in our calculations because that the formation energy of these antisite defects are expected to be high [23].

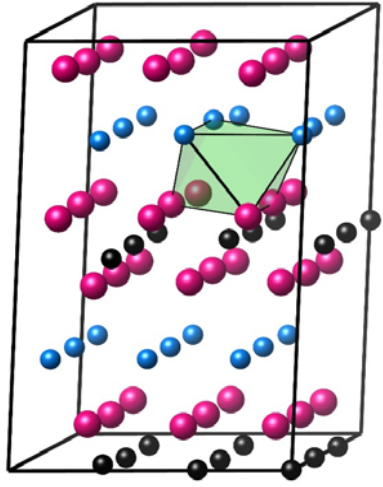

Fig. 8. The interstitial site denoted by the octahedra formed by surrounding Cr and Al atoms.

Table 3. Formation energies (in eV) of on-lattice defects in $Cr_2AlC$ including vacancies, interstitials, antisite defects.

| Defect | Formation energy (eV) |
| --- | --- |
| $V_{Cr}$ | 1.936 |
| $V_{Al}$ | 2.090 |
| $V_C$ | 0.976 |
| $Cr_i$ | 4.526 |
| $Al_i$ | 5.226 |
| $C_i$ | 2.192 |
| $Cr_{Al}$ | 0.982 |
| $Al_{Cr}$ | 1.362 |

The Cr vacancy and Al vacancy have comparable formation energies which are higher than that of C vacancy, indicating that C atoms are easier to be knocked out from their original positions than Cr and Al atoms. For the interstitial type detects, $C_i$ exhibits the lowest formation energy, indicating that $C_i$ is the most stable interstitial in the octahedra located between the Cr layer and Al layer. The formation energy of antisite defect $Al_{Cr}$ is larger than that of antisite defect $Cr_{Al}$. Both antisite defects display much smaller formation energies than those interstitials $Cr_i$ and $Al_i$. Therefore, after being displaced, the Cr and Al atoms would rather become antisite defects than interstitials.

## 5. Discussions

The XRD patterns and SAED patterns showed a structural transition without obvious lattice disorder in the three samples irradiated to doses above 1 dpa. These three irradiated samples have similar lattice parameters, indicating that the modified structure is stable up to the highest dose of 5.2 dpa. Besides, nano-indentation tests revealed comparable hardness values for these three irradiated samples. These phenomena indicate that: (1)He ion irradiation and Xe ion irradiation have the same effects in $Cr_2AlC$ material, and (2) the irradiation effects on the microstructure and mechanical properties of $Cr_2AlC$ material saturate above a certain dose. Similar hardness saturation has been reported for the 95 MeV Xe-ion-irradiated $Ti_3(Si_{0.95}Al_{0.05})C_2$ material after a dose of $1 \times 10^{15}$ ions/cm$^2$ [10]. For 211 phase $Cr_2AlC$, the saturation dose should be no larger than $4 \times 10^{14}$ ions/cm$^2$ for Xe ion irradiation (1 dpa).

Saturation of irradiation effects in $Cr_2AlC$ can be explained based on the mechanism of irradiation-induced antisite defects ($Cr_{Al}$ and $Al_{Cr}$) and C interstitials. The irradiation-induced antisite defects $Cr_{Al}$ and $Al_{Cr}$ leads to a mixture of Cr and Al in each atom layer. As the ion dose increases, the quantity of antisite defects increases until reaching its maximum at a certain dose, after which the antisite defects could be continuously produced by ion irradiation, but the mixture of Al and Cr atoms in each atom layer reaches dynamic equilibrium. The irradiation-induced C interstitials leads to a redistribution of C atoms in all the octahedral holes. After irradiation to a certain dose, the C atoms will randomly occupy the octahedral holes and the interlamellar spacing becomes uniform. The certain dose value for reaching saturation of irradiation effects is not decided in our study, but is estimated to be lower than 1 dpa for $Cr_2AlC$ material.

Based on the formation energies of various defects in $Cr_2AlC$, the displaced Cr and Al atoms would rather become antisite defects than the interstitials, while the displaced C atoms prefer to become interstitials, which support the mechanism of irradiation-induced antisite defects ($Cr_{Al}$ and $Al_{Cr}$) and C interstitials.

The mechanism described above, along with irradiation-induced phase transition (from α phase to β phase), have been used successively to understand the SAED pattern and XRD pattern evolutions after ion irradiation for two 312 phases: $Ti_3SiC_2$ and $Ti_3AlC_2$ [15,16]. Zhao et al. calculated the defects formation energies for these two 312 phases [23]. Take $Ti_3AlC_2$ as an example, the antisite defects $Ti_{Al}$ and $Al_{Ti}$ exhibit lower formation energies than the interstitials $Ti_i$ and $Al_i$, respectively. For the C atoms, the interstitials in the Al layer and interstitials between the Ti layer and Al layer show much lower formation energies than the antisite defects $C_{Ti}$ and $C_{Al}$. These calculations also support the mehcanism described above.

**6. Conclusions**

The microstructural evolution and hardness changes of ion-irradiated $Cr_2AlC$ materials were studied via TEM characterization, XRD analysis and nano-indentation test. Xe ion irradiation and He ion irradiation have the same effects in $Cr_2AlC$. After irradiation, the material underwent a structural transition with an increased *c* lattice parameter and a decreased *a* lattice parameter. The nanolamellar structure was readily destroyed, but the material remained crystalline up to a dose of 5.2 dpa. The three samples irradiated to doses above 1 dpa have comparable lattice parameters and hardness values, suggesting a saturation of irradiation effects in $Cr_2AlC$. The mechanism of irradiation-induced antisite defects ($Cr_{Al}$ and $Al_{Cr}$) and C interstitials was used to explain the structural transition and irradiation effects saturation in $Cr_2AlC$ material. The formation energies of various defects in $Cr_2AlC$ were calculated and support this mechanism.

**Acknowledgments**

This work is supported by The China-Australia Joint Research Project (2014DFG60230), the strategic priority research program of Thorium-based Molten-Salt Reactor (TMSR) (Grant No. XDA02040100), and the National Natural Science Foundation of China (91226202).